\documentclass[reprint, twocolumn,amsmath,amssymb,aps,prd,longbibliography]{revtex4-1}

\usepackage{graphicx}
\usepackage{dcolumn}
\usepackage{bm}
\usepackage[caption=false]{subfig}
\usepackage[usenames, dvipsnames]{color}
\graphicspath{ {Images/} }

\begin{document}

\title{Comparing two models for measuring the neutron star equation of state from gravitational-wave signals}

\author{Matthew F. Carney, Leslie E. Wade, Burke S. Irwin}
\affiliation{%
 Kenyon College,
 Gambier, OH 43022,
 USA
}

\begin{abstract}
Observations of gravitational-wave signals from binary neutron-star mergers, like GW170817, can be used to constrain the neutron-star equation of state (EoS).
One method involves modeling the EoS and measuring the model parameters through Bayesian inference.
A previous study \cite{RecoveringNSEOS} has demonstrated the effectiveness of using a phenomenologically parameterized piecewise polytrope to extract constraining information from a simulated population of binary neutron-star mergers.
Despite its advantages compared to more traditional methods of measuring the tidal deformability of neutron stars, notable deficiencies arise when using this EoS model.
In this work, we describe in detail the implementation of a model built from a spectral decomposition of the adiabatic index that was used by the LIGO-Virgo Collaboration in Ref.~\cite{LIGO_EOS} to constrain the neutron star EoS from GW170817.
We demonstrate its overall consistency in recovering the neutron star EoS from a simulated signal to the piecewise polytropic implementation used in Ref.~\cite{RecoveringNSEOS} and explain any differences that arise.
We find that both models recover consistent tidal information from the simulate signals with tightest constraints on the EoS around twice nuclear saturation density.
As expected, the statistical error that arises in the piecewise polytropic representation near the fixed joining densities is greatly reduced by using the spectral model.
In addition, we find that our choice of prior can have a dominant effect on EoS constraints.
\end{abstract}

\maketitle

\section{\label{sec:intro}Introduction}
Gravitational-wave signals from coalescing binary neutron-star (BNS) systems offer a new way to probe the physics of nuclear matter in the cold, high-density region of its equation of state (EoS) accessible only by neutron stars.
An EoS is a relationship between state variables, such as pressure and density, of matter.
Though the nuclear EoS has remained largely unconstrained at supranuclear densities \cite{NSMassRadius}, future electromagnetic missions such as NICER \cite{NICER} promise to tighten our understanding of the exotic states of matter found in neutron stars.
Still, gravitational waves will provide essential, independent inference to compliment and enhance any electromagnetic measaurements.
Indeed, the LIGO-Virgo Collaboration has already used the first BNS gravitational-wave signal GW170817 to place constraints on the tidal deformability of the two stars before merger \cite{FirstBNS,aasi2015advanced,acernese2014advanced}.
Measurements of both the mass and tidal deformability of the system's neutron stars lead directly to constraints on the neutron-star EoS, as several followup studies have demonstrated \cite{De:2018uhw,margalit2017constraining,bauswein2017neutron,zhou2018constraints,rezzolla2018using,fattoyev2018neutron,nandi2018hybrid,paschalidis2018implications,ruiz2018gw170817,annala2018gravitational,raithel2018tidal,most2018new}.

With more gravitational-wave detections assuredly on the way, it is vital to establish sophisticated yet tractable methods for extracting the neutron-star EoS from the multi-source analysis of future BNS signals.
Modeling the neutron-star EoS instead of mapping macroscopic properties of the neutron stars, like the masses and tidal deformability or radius, to the EoS has distinct advantages in this regime \cite{RecoveringNSEOS,riley2018parametrised,raaijmakers2018pitfall,abdelsalhin2018solving}.
For instance, operating under the assumption that the neutron-star equation of state is universal \cite{glendenning2012compact}, each detection will yield measurements of the same relationship, which makes combining information from each subsequent system straightforward.
Additionally, the Bayesian prior is defined on the EoS itself, as opposed to any exterior parameters like the tidal deformability or radius.
Lastly, any additional physical or observational EoS information can be easily folded into the analysis through the prior.

A previous study \cite{RecoveringNSEOS} has demonstrated the effectiveness of measuring the parameters of a phenomenologically motivated piecewise polytropic parameterization of the neutron-star EoS \cite{4Piece} from a population of simulated BNS gravitational-wave events.
Increased statistical error in the recovered EoS arose at the fixed joining densities of the piecewise model, motivating a different choice of parameterization.
In this work, we detail the implementation of an EoS model built from a spectral decomposition of the adiabatic index \cite{LindblomSpectral} chosen because it better matches a wide variety of candidate EoSs than piecewise polytrope models with the same number of parameters, as shown in Refs.~\cite{LindblomSpectral,causalspectral}.
We compare these two implementations and find that the model based on a spectral decomposition of the adiabatic index alleviates the increased statistical error in the density regions corresponding to the fixed stitching points of the piecewise polytrope model.
We also touch on the important effect of one's choice of prior when performing EoS inference.

This paper is organized as follows: In Sec.~\ref{sec:bayesian}, we outline the Bayesian parameter estimation techniques used to infer the source parameters of a gravitational-wave event.
In Sec.~\ref{sec:settings}, we present the waveform models and settings used in our study.
In Sec.~\ref{sec:4piece} and ~\ref{sec:spectral}, we discuss the two EoS models compared.
Finally, in Sec.~\ref{sec:results} and ~\ref{sec:conclusion}, we discuss the results of the comparison and future work.

\section{\label{sec:methods} Methods}

\subsection{\label{sec:bayesian}Bayesian Inference}

To compare the utility of the two parameterized models for measuring the EoS, we seek to estimate the posterior probability density function (PDF) $p(\boldsymbol{\theta}|d)$, where $d$ represents the LIGO and Virgo data, and $\boldsymbol{\theta}$ represents the source parameters of the emitting system.
Bayes' Theorem relates the posterior PDF to the prior PDF $p(\boldsymbol{\theta})$ and the likelihood $p(d|\boldsymbol{\theta})$:
\begin{eqnarray}
p(\boldsymbol{\theta}|d) &=& \frac{p(d|\boldsymbol{\theta})p(\boldsymbol{\theta})}{p(d)}\\
&\propto& p(d|\boldsymbol{\theta})p(\boldsymbol{\theta}).
\label{Bayes}
\end{eqnarray}
The evidence $p(d)$ is effectively a normalization factor to satisfy the condition that the posterior density function must integrate to one.
The prior PDF represents any {\it a priori} knowledge about the source parameters. The likelihood is the probability that the data $d$ is reproduced by a system with source parameters $\boldsymbol{\theta}$.
The single-detector likelihood takes the functional form
\begin{equation}
p(d|\boldsymbol{\theta}) \propto \exp \left [ -2 \int_{0}^{\infty} \frac{|d(f) - h(\boldsymbol{\theta},f)|^2}{S_{n}(f)} df\right ],
\label{likelihood}
\end{equation}
assuming stationary, Gaussian noise, where $S_n(f)$ is the one-sided power spectral density (PSD) of the noise, $d(f)$ is the Fourier transform of the data, and $h(\boldsymbol{\theta},f)$ is a gravitational waveform model in the frequency domain.
Our analysis includes simulated data from three detectors, the two Advanced LIGO detectors and the Advanced Virgo detector, for which the network likelihood is just the product of the single-detector likelihoods.

The direct calculation of the posterior over a large parameter space is formidable.
For a typical gravitational waveform model, the dimensionality of the parameter space can exceed fifteen.
We therefore rely on sampling techniques like the Markov-chain Monte Carlo (MCMC) algorithm, which draws samples from the underlying posterior PDF.
Specifically, we use the LALInference gravitational-wave parameter estimation software package \cite{lalinference} found in the open-source LIGO Algorithm Library (LAL) for our analysis \cite{lalsuite}.

\subsection{\label{sec:settings}Waveform, injection, and settings}

We use the TaylorF2 waveform approximant with leading order and next-to-leading order tidal corrections to the phase (see Ref.~\cite{WadeSystematics} and references therein) as the waveform templates used in the likelihood calculations.
These tidal corrections depend on just two parameters: the tidal deformability $\Lambda=(2/3)k_2[c^2R / (G m)]^5$ of each star, where $k_2$ is the second Love number, $R$ is the star's radius and $m$ is its mass.
The inspiral of the waveform is computed from a frequency of 30~Hz to the frequency at the innermost stable circular orbit, $f_{\text{isco}}=c^3/(6^{3/2}\pi G M)$, of a test particle about a Schwarzschild black hole of mass $M$.
Though neutron stars with stiff EoSs may merge before $f_{\text{isco}}$, this was found to affect EoS measurability by less than $5\%$ \cite{WadeSystematics} and was ignored in this study.
Consistent with Refs.~\cite{RecoveringNSEOS} and \cite{WadeSystematics}, we inject a simulated signal into zero noise.
However, we still incorporate detector noise in our analysis through the PSDs included in our likelihood calculations.
This approach insulates our study from the statistical fluctuations of individual noise realizations while preserving the overall effect of the noise \cite{BasicGWPE,nissanke2010exploring}.
We use the zero-detuned high power configuration for the Advanced LIGO PSDs \cite{ligo-zdhp} and an Advanced Virgo PSD based on Eq.~6 of Ref.~\cite{manzotti2012prospects}.
Additionally, we also use the tidally-corrected TaylorF2 waveform for our injected signal to avoid the systematic error associated with waveform modeling \cite{WadeSystematics,RecoveringNSEOS}.

The strength of the signal relative to the noise is given by the signal-to-noise ratio (SNR), which for a single detector is defined to be
\begin{equation}
\text{SNR}_n = \sqrt{4\int_0^{\infty}\frac{|h(\boldsymbol{\theta},f)|^2}{S_n(f)} df}.
\end{equation}
For a network of $n$ detectors, the SNR is
\begin{equation}
\text{SNR}_{\text{net}} = \sqrt{\sum_n \rho_n^2}.
\end{equation}

For this study, we perform our analysis on just the highest-SNR event from the baseline BNS population analyzed in Ref.~\cite{RecoveringNSEOS}.
This corresponds to a 1.52-1.52~$\text{M}_{\odot}$, non-spinning BNS system with the MPA1 EoS \cite{muther1987nuclear,NSMassRadius} and $\text{SNR}_{\text{net}}=64$.
The EoS is recovered using two different parameterizations: a four-parameter piecewise polytrope and a four-parameter spectral decomposition of the adiabatic index, discussed in sections \ref{sec:4piece} and \ref{sec:spectral}, respectively.

\subsection{\label{sec:4piece}Piecewise polytrope}

The piecewise polytrope parameterization of the EoS is constructed by stitching together three individual polytropes in the high-density portion of the EoS and anchoring to a fixed low-density EoS~\cite{4Piece}.
Expressed as a relationship between the pressure and density of the neutron star, the piecewise polytrope takes the form
\begin{equation}
p(\rho) = K_{i}\rho^{\Gamma_{i}},
\end{equation}
where $p$ is the pressure, $\rho$ is the baryon mass density, $K_{i}$ is a constant of proportionality, $\Gamma_{i}$ is the adiabatic index, and $i$ labels the segment of the piecewise function.
The model requires four parameters: $\log(p_1)$, $\Gamma_1$, $\Gamma_2$, and $\Gamma_3$.
The model is constructed by dropping a polytrope with adiabatic index $\Gamma_1$ anchored at pressure $p_1$ and a fixed joining density $\rho_1=10^{14.7}$~g~cm$^{-3}\approx1.8\,\rho_{\rm nuc}$, where the nuclear saturation density $\rho_{\rm nuc}\sim2.8\times10^{14}$~g~cm$^{-3}$, down to the fixed low-density EoS. 
A second polytrope with adiabatic index $\Gamma_2$ is then attached from $(\rho_1,p_1)$ to a fixed joining density $\rho_2=10^{15.0}$~g~cm$^{-3}\approx3.6\,\rho_{\rm nuc}$.
Finally, the EoS is completed by attaching a third polytrope with adiabatic index $\Gamma_3$ at $\rho_2$.
This form of the EoS has been shown to reproduce macroscopic observables for a wide range of candidate EoSs to within a few percent with just four free parameters \cite{4Piece}, motivating its use in a gravitational-wave parameter estimation context.

For our analysis, we assume a uniform prior on all EoS parameters, asserting that $\log (p_1) \in [33.6,35.4]$, $\Gamma_1 \in [2.0,4.5]$, and $\Gamma_2,\Gamma_3 \in [1.1,4.5]$ for all EoS samples, consistent with Ref.~\cite{RecoveringNSEOS}.
The flat prior ranges are chosen to encompass a wide range of candidate EoSs and impose thermal stability ($d\epsilon/dp > 0$). In addition to uniform priors on each parameter, we require the following of all EoS samples:
\begin{enumerate}
\item Causality: the speed of sound $v_s = \sqrt{dp/d\epsilon}$ must be less than the speed of light up to the central pressure of the heaviest neutron star supported by the EoS. 
\item Observational consistency: the EoS has a maximum mass above observed neutron star masses.
\item The components are neutron stars: the mass of each component is supported by the same EoS.
\end{enumerate}
Since the piecewise polytropic parameterization reproduces macroscopic observables to within a few percent, in practice we only enforce the causality prior when $v_s>1.1\,c$ to allow for the possibility of causal EoSs being fit by acausal polytropic representations.
When imposing the observational prior, we require that all EoS samples support a maximum neutron-star mass of 1.97\,$M_{\odot}$, which corresponds to the $1\sigma$ lower bound of the mass measurement of pulsar PSR J0348+0432 in Ref.~\cite{antoniadis2013massive}.
If any of these conditions are not met, the prior value is set to 0 and the set of parameters is rejected by the MCMC sampler. 

We used LALSimulation's preexisting support for the piecewise polytrope model and modified LALInference to sample in the model's parameters \cite{lalsuite}, which is an intentional deviation from the two-stage approach of Ref.~\cite{RecoveringNSEOS}.
The EoS parameters $(\log(p_1),\Gamma_1,\Gamma_2,\Gamma_3)$ and masses $(m_1,m_2)$ of each sample are mapped to the tidal deformabilities $\Lambda_i(m_i;\log(p_1),\Gamma_1,\Gamma_2,\Gamma_3)$ of each star via an integration of the Tolman-Oppenhimer-Volkoff (TOV) equations~\cite{OVeqs}.
The two tidal deformability parameters $(\Lambda_1,\Lambda_2)$ are then used to compute the gravitational waveform.
Although the execution of this approach varies slightly from the methods described in Ref.~\cite{RecoveringNSEOS}, we found that it produces entirely consistent results.

\subsection{\label{sec:spectral}Spectral decomposition}
The adiabatic index of the EoS can be spectrally decomposed onto a set of polynomial basis functions,
\begin{equation}
\Gamma(x) = \exp\left(\sum_{k} \gamma_k x^k\right),
\end{equation}
where $\gamma_k$ are the expansion coefficients, and $x=\log(p/p_0)$ is a dimensionless pressure variable taken with respect to some reference pressure, $p_0$~\cite{LindblomSpectral}. Since the basis functions are differentiable, the discontinuities in the derivative of the EoS that are present in the polytropic representation are absent in this model. From the adiabatic index, the EoS can be constructed in the form of energy density as a function of pressure by direct integration of
\begin{equation}
\frac{d\epsilon}{dp} = \frac{\epsilon+p}{p\Gamma(p)},
\end{equation}
which can be reduced to quadratures:
\begin{equation}
\epsilon(p) = \frac{\epsilon_0}{\mu(p)} + \frac{1}{\mu(p)}\int_{p_0}^{p} \frac{\mu(p')}{\Gamma(p')}dp',
\label{eofpspectral}
\end{equation}
where $\mu(p)$ is
\begin{equation}
\mu(p) = \exp \left (-\int_{p_0}^{p} \frac{dp'}{p' \Gamma(p')} \right ).
\label{muofpspectral}
\end{equation}
Because the adiabatic index is decomposed onto polynomials, Eqs.~\ref{eofpspectral} and \ref{muofpspectral} can be calculated to double precision via Gaussian quadrature numerical integration with just ten evaluation points per integral.
The spectral parameterization has smaller residuals than the piecewise polytrope with fewer parameters when fitting to a wide range of candidate EoSs~\cite{LindblomSpectral}.
Even in cases in which the candidate EoSs contain phase transitions, the spectral fit with four parameters is about as accurate as the piecewise polytrope model. 

We choose to use a four-parameter spectral representation for this comparison, since the piecewise polytrope model is built on four parameters.
Just as with the polytrope model, we impose uniform priors on all of the spectral model parameters: $\gamma_0 \in [0.2,2.0]$, $\gamma_1 \in [-1.6,1.7]$, $\gamma_2 \in [-0.6,0.6]$, and $\gamma_3 \in [-0.02,0.02]$.
We assume that the system is a BNS and enforce the causality and observational priors outline in Sec.~\ref{sec:4piece}.
We again use $v_s>1.1\,c$ for our causality constraint to allow acuasal fits of causal EoSs.
Finally, we also require that the adiabatic index be confined to $\Gamma(p)\in[0.6,4.5]$.
These priors were chosen to encompass a wide variety of candidate EoSs and to explore at least as much of the EoS parameter space as the polytrope model.

The high density EoS is stitched to a low-density EoS at $p_0$.
We chose the SLy EoS of Ref.~\cite{SLyEOS} for our low-density EoS.
We also made two distinct choices for where to start the spectral EoS representation, and we present both choices here.
We first chose $p_0\approx5.3\times10^{30}$~dyne~cm$^{-2}$, which is just below the lowest density that the polytrope model attaches to its low-density EoS within our chosen prior range on the parameters.
This choice was made in order to fully encompass the EoS parameter space accessible to the polytrope model for this comparison study.
This choice effectively resulted in an extremely conservative prior on the low-density portion of the spectral EoS, which is discussed in Sec.~\ref{sec:results}.
However, since densities below $0.5\,\rho_{\rm nuc}$ do not significantly alter the macroscopic observables of neutron stars \cite{lattimer2001neutron,raithel2016neutron}, we next chose $p_0\approx5.4\times10^{32}$~dyne~cm$^{-2}$, which corresponds to a density just below $0.5\,\rho_{\rm nuc}$ on our low-density EoS.
This choice reflects that the nuclear EoS is relatively well-constrained below $\rho_{\rm nuc}$ \cite{lattimer2012nuclear}.

We added this spectral decomposition model to the LALSimulation software package and modified LALInference to sample in the model's parameters \cite{lalsuite}.
The EoS parameters $(\gamma_0,\gamma_1,\gamma_2,\gamma_3)$ and masses $(m_1,m_2)$ of each sample are mapped to the tidal deformabilities $\Lambda_i(m_i;\gamma_0,\gamma_1,\gamma_2,\gamma_3)$ of each star via an integration of the TOV equations.
The two tidal deformability parameters $(\Lambda_1,\Lambda_2)$ are then used to compute the gravitational waveform.

\section{\label{sec:results}Results}

\begin{figure}[t]
\centering
\includegraphics[width=0.5\textwidth]{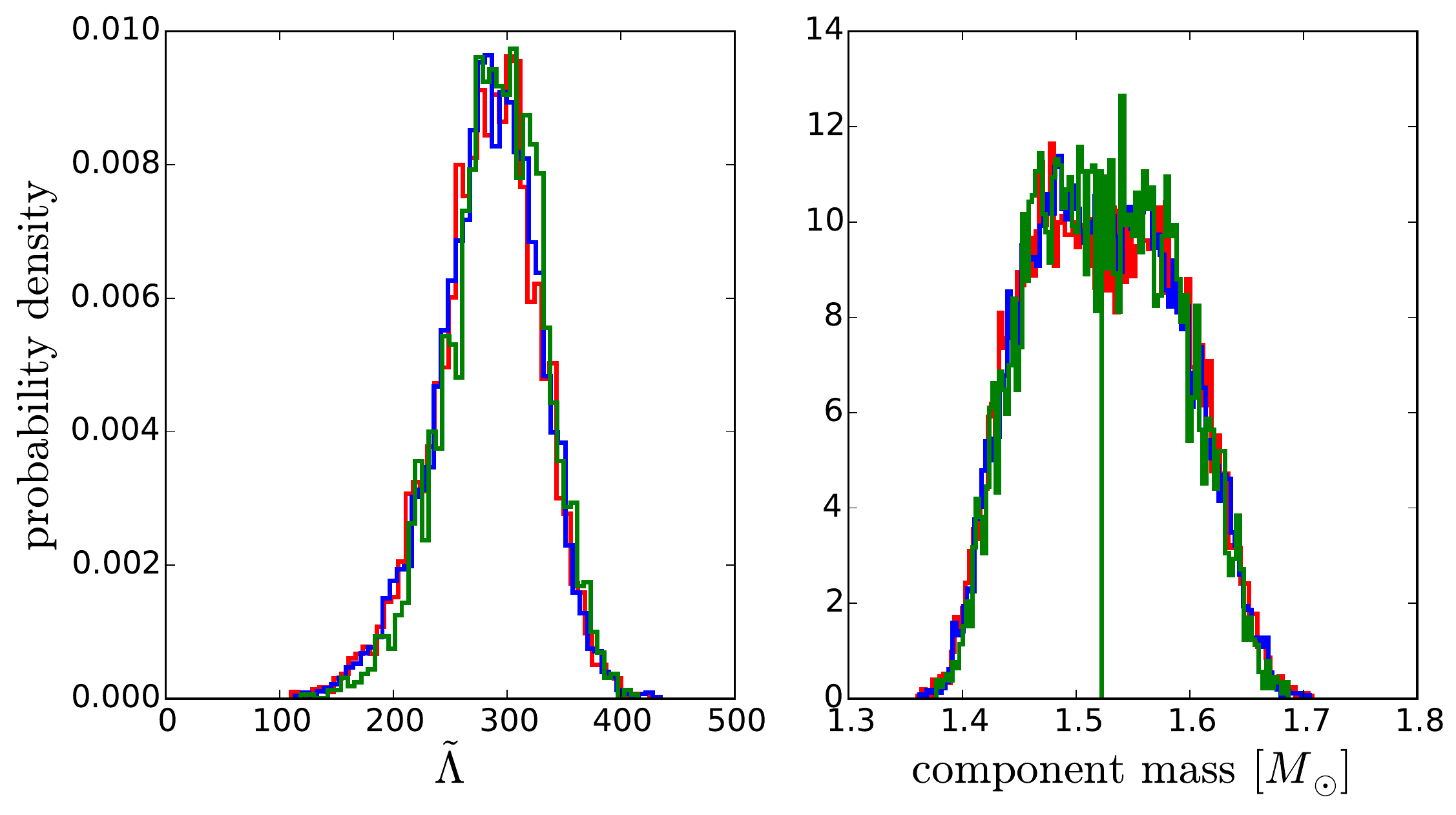}
\caption{1D posterior PDFs on the measurable tidal information $\tilde{\Lambda}$ (left panel) and the two component masses $m_1$ and $m_2$ (right panel) of the simulated BNS signal.
The polytrope model (red), the low-$p_0$ spectral model (blue), and the high-$p_0$ spectral model (green) recover similar PDFs indicating that each extract the same EoS-dependent information from the signal.
Note that for the component mass PDFs, we take $m_1>m_2$ and plot both PDFs for a given model in the same color.}
\label{L-m}
\end{figure}

As discussed in Sec.~\ref{sec:settings}, we perform our comparison on the highest-SNR event from the baseline BNS population of Ref.~\cite{RecoveringNSEOS}.
For simplicity, in this section we refer to the piecewise polytrope model outlined in Sec.~\ref{sec:4piece} as simply the {\it polytrope model}; we refer to the spectral decomposition model outlined in Sec.~\ref{sec:spectral} stitched to the low-density EoS at $p_0\approx5.3\times10^{30}$~dyne~cm$^{-2}$ as the {\it low-$p_0$ spectral model}; and we refer to the spectral decomposition model stitched to the low-density EoS at $p_0\approx5.4\times10^{32}$~dyne~cm$^{-2}$ as the {\it high-$p_0$ spectral model}.

In Fig.~\ref{L-m}, we present 1D marginalized histograms of the component masses $m_1$ and $m_2$ and $\tilde{\Lambda}$, which is the dimensionless version of $\tilde{\lambda}$ from Ref.~\cite{flanagan2008constraining} and defined to be \cite{favata2014systematic}
\begin{eqnarray}
\nonumber
\tilde{\Lambda} &=& \frac{8}{13}\left[ (1+7\eta-31\eta^2)(\Lambda_1+\Lambda_2) \phantom{\sqrt{4}}\right.\\
&&\left.+\sqrt{1-4\eta}\,(1+9\eta-11\eta^2)(\Lambda_1-\Lambda_2)\right],
\end{eqnarray}
where $\eta=m_1 m_2/(m_1+m_2)^2$.  $\tilde\Lambda$ is the contribution to the compact binary coalescence waveform of the individual tidal deformabilities $\Lambda_1$ and $\Lambda_2$ that is measurable with advanced ground-based interferometers \cite{WadeSystematics}.
Throughout we assume that $m_1>m_2$.
Fig.~\ref{L-m} shows that all three models, the polytrope model (red), the low-$p_0$ spectral model (blue), and the high-$p_0$ spectral model (green), recover nearly identical marginalized PDFs on the component masses and the measurable tidal information, $\tilde{\Lambda}$, of the system.
This demonstrates that each model is extracting the same information from the signal, and any difference between the EoS constraints from any of the three models is only due to the mapping from this information onto the EoS and differences in the incorporated priors.

\begin{figure}[!t]
\centering
\includegraphics[width=\textwidth]{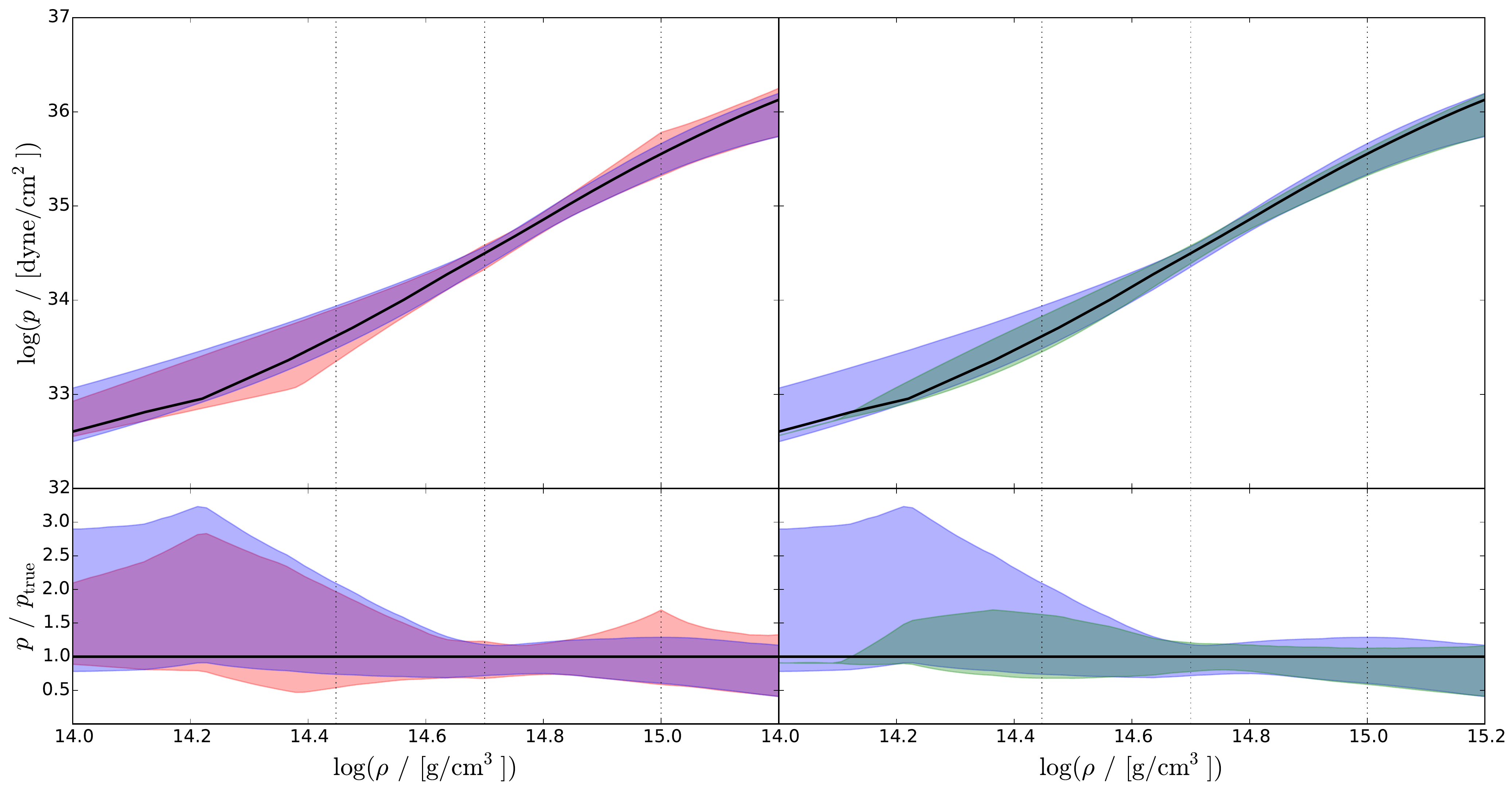}
\caption{Left: 90\% credible regions of the recovered EoS using the polytrope model (red) and the low-$p_0$ spectral model (blue) in pressure-density space. Right: 90\% credible regions of the recovered EoS using the low-$p_0$ spectral model (blue) and the high-$p_0$ spectral model (green) in pressure-density space.
The bottom panels divide the pressure values at each density by the pressure of the injected EoS at each density.
The vertical dotted lines label $\rho_{\rm nuc}$, $\rho_1\approx1.8\,\rho_{\rm nuc}$, and $\rho_2\approx3.6\,\rho_{\rm nuc}$, from left to right, where $\rho_1$ and $\rho_2$ are the joining densities for the polytrope model.
The injected EoS, MPA1, is represented in each panel by a solid, black line.  For details regarding the construction of these plots, see Sec.~\ref{sec:results}.  Each model achieves a consistent measurement near $2\,\rho_{\rm nuc}$.
The differences at low-densities can be attributed to the choice of prior but are included for comparison to Ref.~\cite{RecoveringNSEOS}.
Notice, however, in the bottom left panel that the increased width in the credible intervals experienced by the polytrope model near its joining densities are mitigated by the smoother spectral model.}
\label{p-rho}
\end{figure}

In Fig.~\ref{p-rho}, we present these constraints on $p(\rho)$ using each EoS parameterization.
To make these plots, the credible regions are determined by first discretizing the density space in Fig.~\ref{p-rho}, which yields a set of density points $\rho_i$.
We then compute $p(\rho_i)$ for every EoS sample and histogram the resulting pressure distribution for each density.
Finally, the 90\% credible region is computed for each histogram and stitched together to bound the credible region in pressure-density space.
The $p/p_{\rm true}$ panels divide each pressure value by the pressure value of the injected EoS (MPA1) at each considered density.
In the left panels of Fig.~\ref{p-rho}, we overlay the 90\% credible regions from the polytrope model (red) and the low-$p_0$ spectral model (blue); and in the right panels, we overlay the 90\% credible regions from the low-$p_0$ spectral model (blue) and the high-$p_0$ spectral model (green).

\begin{figure}[!t]
\centering
\includegraphics[width=\textwidth]{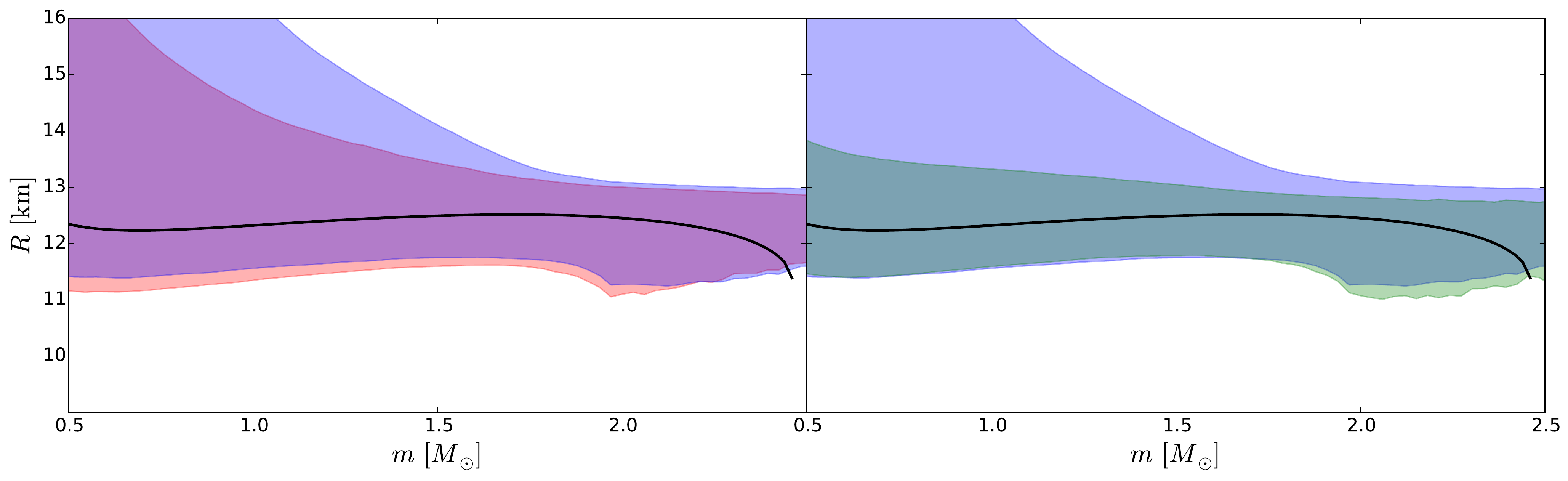}
\caption{Left: 90\% credible regions of the recovered EoS using the polytrope model (red) and the low-$p_0$ spectral model (blue) in radius-mass space.
Right: 90\% credible regions of the recovered EoS using the low-$p_0$ spectral model (blue) and the high-$p_0$ spectral model (green) in radius-mass space.
The injected EoS, MPA1, is represented in both panels by a solid, black line.
For details regarding the construction of these plots, see Sec.~\ref{sec:results}.
The differences at low mass can be attributed to the choice of prior.}
\label{R-M}
\end{figure}

Since the EoS is often mapped to mass-radius space when considering neutron stars, we also present this mapping in Fig.~\ref{R-M}.
In just the same way that we produce $p(\rho)$ credible intervals, we discretize the mass space in Fig.~\ref{p-rho}, which yields a set of points $m_i$.
We then compute $R(m_i)$ by solving the TOV equations for every EoS sample and histogram the resulting radius distribution for each mass.
If the EoS sample does not support a neutron star at mass $m_i$, it is excluded from the histogram, which accounts for the credible region turnover above 1.97\,$M_\odot$.
Finally, the 90\% credible region is computed for each histogram and stitched together to bound the credible region in radius-mass space.
In the left panel of Fig.~\ref{R-M}, we overlay the 90\% credible regions from the polytrope model (red) and the low-$p_0$ spectral model (blue); and in the right panel, we overlay the 90\% credible regions from the low-$p_0$ spectral model (blue) and the high-$p_0$ spectral model (green).

Each credible region in Fig.~\ref{p-rho} is narrowest just above $\rho=10^{14.7}$~g~cm$^{-3}$, which is the first fixed joining density of the polytrope model.
However, there is a noticeable widening of the credible regions near the piecewise polytrope fixed joining densities, consistent with the findings of Ref.~\cite{RecoveringNSEOS}, which is mitigated by using the spectral model.
The ostensibly improved constraint at low densities with the polytrope model compared to the low-$p_0$ spectral model (see left panels of Fig.~\ref{p-rho}) is an artifact of the overly-conservative prior on the low-$p_0$ spectral model.
This is made clear by comparing the low-density credible regions of these two models against those of the high-$p_0$ spectral model (see right panels of Fig.~\ref{p-rho}).
This feature also manifests itself in radius-mass space, shown in Fig. \ref{R-M}, where it appears as though the polytrope model results in tighter constraints than the spectral model.
However, again comparing these constraints against the high-$p_0$ spectral model demonstrates that this is merely a difference in prior.
If we applied similarly tight priors to the low-density portion of the polytrope EoS model, it too would see tighter credible intervals in these regions.

\section{\label{sec:conclusion}Conclusions and Outlook}

In this work, we used Bayesian inference to extract the EoS information inherent in a simulated gravitational-wave signal from a binary neutron-star system.
We accomplished this by modeling the EoS and measuring the model parameters.
We used two different EoS parameterizations for this study: a piecewise polytrope and a spectral decomposition of the adiabatic index.
We successfully implemented both models into LIGO's flagship parameter estimation software package LALInference~\cite{lalinference}, which is an improvement to the method outlined in Ref.~\cite{RecoveringNSEOS}, and used this software to analyze a simulated high-SNR BNS signal.
We then compared the recovered posterior distributions using each model and presented our main findings in Figs.~\ref{p-rho} and \ref{R-M}.

Figs.~\ref{p-rho} and \ref{R-M} show that the spectral implementation recovers EoS constraints that are very consistent with the implementation of the polytrope model first published in Ref.~\cite{RecoveringNSEOS} and improved and reproduced here.
Notice that the measurement of the EoS is happening roughly around $\rho\approx2\,\rho_{\rm nuc}$ for both models.
However, the spectral model mitigates the widening of the 90\% credible intervals near the piecewise polytrope joining densities.
Note that this does not indicate an inherent advantage over piecewise polytrope models.
These credible interval widenings are a direct result of the fixed joining densities of the polytrope implementation and could likewise have been mitigated by using a piecewise polytrope model with fewer pieces and parameterized joining densities.
Implementing such a model is a subject of future work.
Lastly, it is clear that choosing an appropriate prior has a noticeable impact on the credible intervals in these spaces.
Modeling the EoS is an effective way to connect the measurements of the EoS around $2\,\rho_{\rm nuc}$ from gravitational-wave observations to low-density EoS constraints made in the laboratory.
In making these measurements, more important than the EoS model used is the choice of prior. 

Here, we reconstructed the neutron-star EoS from just a single simulated BNS detection.
However, this approach naturally allows for the inclusion of any number of neutron star detections \cite{RecoveringNSEOS}.
In particular, systems with distinctive source parameters will constrain slightly different regions of the EoS.
For example, a high-mass BNS system could contain complementary EoS information to a system with smaller masses.
Thus, combining information from multiple observations would result in tighter constraints on the equation of state as a whole~\cite{del2013demonstrating,agathos2015constraining,RecoveringNSEOS}.

The flexibility of the spectral parameterization allows for some interesting future developments.
Including more degrees of freedom in the model may compromise computational efficiency for higher accuracy if the EoS being recovered has sharp curvature.
Since the spectral model is robust to a higher number of parameters, this could be studied by performing the same analysis discussed here with a higher or lower number of expansion parameters and comparing Baye's factors to determine an optimal number to use.
The possibility also exists that there is no fixed optimal value for the number of terms to include in the decomposition.
A potential supplement to the parameter estimation routine could be to include the number of expansion coefficients as a parameter itself.
This would allow the data to inform the optimal number of terms to include and simultaneously determine their values, similar to methods described in Ref.~\cite{Green}.
A similar analysis could be performed for a more generic piecewise polytrope model, if implemented.

Recently, a modification to the spectral model explored in this work has been proposed in which the causality constraint is self-imposed \cite{causalspectral}.
This elegant addition to the model would lead to improved computational efficiency. Implementing the modified spectral model will also be a subject of future work.

Lastly, each EoS model with some finite number of parameters will presumably never perfectly fit the true EoS.
In future work, we will quantify this modeling error on a large set of candidate EoSs so that we can marginalize over this error, similar to what is done in Ref.~\cite{chatziioannou2018measuring}.

\section{Acknowledgements}

We sincerely thank Tyson Littenberg for many helpful insights and suggestions, Jocelyn Read and Richard O'Shaughnessy for several helpful discussions, Lee Lindblom for advice in implementing the spectral model, Benjamin Owen for comments on this manuscript, and Madeline Wade for extensive technical assistance.
This work was supported in part by NSF Grant No. PHY-1607178 and the Kenyon College Summer Science Scholars program.

\onecolumngrid

\twocolumngrid

\bibliography{mybib}

\end{document}